\documentclass[aps,prl,twocolumn,groupedaddress,showpacs]{revtex4}

\usepackage{graphicx}
\usepackage{amssymb}
\usepackage{amsmath}

\begin{document}
\preprint{LLNL-JRNL-402788}

\title{{\em Ab Initio} Many-Body Calculations of $\boldsymbol{n}$-$^{\boldsymbol{3}}$H, $\boldsymbol{n}$-$^{\boldsymbol{4}}$He, 
$\boldsymbol{p}$-$^{\boldsymbol{3,4}}$He, and $\boldsymbol{n}$-$^{\boldsymbol{10}}$Be Scattering}


\author{Sofia Quaglioni}
\email[]{quaglioni1@llnl.gov}
\author{Petr Navr{\'a}til}
\email[]{navratil1@llnl.gov}
\affiliation{Lawrence Livermore National Laboratory, P.O. Box 808, L-414, Livermore, CA 94551, USA}


\date{\today}

\begin{abstract}
We develop a new {\em ab initio} many-body approach capable 
of describing simultaneously both bound and scattering states in light nuclei, 
by combining the resonating-group method with the use of realistic interactions, and a microscopic and consistent description
of the nucleon clusters. This approach preserves translational symmetry and Pauli principle.
We present phase shifts for neutron scattering on $^3$H, $^4$He and $^{10}$Be and proton
scattering on $^{3,4}$He, using realistic nucleon-nucleon potentials.
Our $A=4$ scattering results are compared to earlier {\em ab initio} calculations. We demonstrate that a proper treatment of the coupling to the $n\,$-${}^{10}$Be continuum is essential to explain the parity-inverted ground state in $^{11}$Be.
\end{abstract}


\maketitle
The development of a quantitative microscopic theory of low-energy nuclear reactions on light nuclei is key to further refine our understanding of the fundamental nuclear interactions among the constituent nucleons, providing, at the same time, accurate predictions of crucial reaction rates for nuclear astrophysics. The enormous difficulties involved with such a project are immediately apparent if one considers the nearly-total lack of  {\em ab initio} calculations for scattering processes involving more than four nucleons overall~\cite{GFMC_nHe4}. Nevertheless, a breakthrough is within reach today thanks to the significant progress achieved in the {\em ab initio} description of the structure of light nuclei and the advent of modern high-performance computers. 
In this Letter we combine a microscopic-cluster technique,  the resonating-group method (RGM)~\cite{RGM}, and a very successful structure approach, 
the {\em ab initio} no-core shell model (NCSM)~\cite{NCSMC12}, into a new microscopic theory ({\em ab initio} NCSM/RGM) capable of treating bound and scattering states of light nuclei in a unified formalism, starting from the fundamental inter-nucleon interactions. Within this new approach we study the $n\,$-${}^3$H, $n\,$-${}^4$He, $n\,$-${}^{10}$Be, and $p\,$-${}^{3,4}$He scattering processes,  and address the parity inversion  of  the $^{11}$Be  ground state (g.s.), using realistic nucleon-nucleon (NN) potentials.

We start from the wave function for a scattering process involving pairs of nuclei that can be cast in the form 
\begin{equation}
|\Psi^{J^\pi T}\rangle = \sum_{\nu} \int dr \,r^2\frac{g^{J^\pi T}_\nu(r)}{r}\,\hat{\mathcal A}_{\nu}\,|\Phi^{J^\pi T}_{\nu r}\rangle\,, \label{trial}
\end{equation}
through an expansion over  binary-cluster channel-states of total angular momentum $J$, parity $\pi$, and isospin $T$,
\begin{eqnarray}
|\Phi^{J^\pi T}_{\nu r}\rangle &=& \Big [ \big ( \left|A-a\, \alpha_1 I_1^{\,\pi_1} T_1\right\rangle \left |a\,\alpha_2 I_2^{\,\pi_2} T_2\right\rangle\big ) ^{(s T)}\nonumber\\
&&\times\,Y_{\ell}\left(\hat r_{A-a,a}\right)\Big ]^{(J^\pi T)}\,\frac{\delta(r-r_{A-a,a})}{rr_{A-a,a}}\,.\label{basis}
\end{eqnarray}
The wave functions of the ($A-a$)- and $a$-nucleon clusters are antisymmetric under exchange of internal nucleons, and depend on translationally-invariant internal coordinates. They are eigenstates of the $H_{(A-a)}$ and $H_{(a)}$ intrinsic Hamiltonians with spin,  parity, isopsin and additional quantum numbers $I_i, \pi_i, T_i$, and $\alpha_i$, respectively, where $i=1,2$. The clusters centers of mass are separated by the relative vector $\vec r_{A-a,a} $.  
Relative angular momentum and channel spin are denoted by $\ell$ and $s$, respectively. The inter-cluster anti-symmetrizer for the $(A-a,a)$ partition in Eq.~(\ref{trial}) can be schematically written as $\hat{\mathcal A}_{\nu}=[(A-a)!a!/A!]^{1/2}\sum_{P}(-)^pP$, where  $P$ are permutations among nucleons pertaining to different clusters, and $p$ the number of interchanges characterizing them. The coefficients of the expansion with respect to the channel index $\nu=\{A-a\,\alpha_1I_1^{\,\pi_1} T_1;\, a\, \alpha_2 I_2^{\,\pi_2} T_2;\, s\ell\}$ are the relative-motion wave functions $g^{J^\pi T}_\nu(r)$, which represent the unknowns of the problem. They can be determined by solving the many-body Schr\"odinger equation in the Hilbert space spanned by the basis states $\hat{\mathcal A}_{\nu}\,|\Phi^{J^\pi T}_{\nu r}\rangle$, 
\begin{equation}
\sum_{\nu}\int dr \,r\,{\mathcal K}^{J^\pi T}_{\nu^\prime\nu}(r^\prime, r)\,g^{J^\pi T}_\nu(r) = 0\,,\label{RGMeq}
\end{equation}
where the integration kernel is given by:
\begin{equation}
{\mathcal K}^{J^\pi T}_{\nu^\prime\nu}(r^\prime, r) = \left\langle\Phi^{J^\pi T}_{\nu^\prime r^\prime}\right|\hat{\mathcal A}_{\nu^\prime}(H-E)\hat{\mathcal A}_{\nu}\left|\Phi^{J^\pi T}_{\nu r}\right\rangle\,.\label{K-kernel} 
\end{equation}
Here $E$ is the total energy in the center-of-mass (c.m.) frame, and $H$ is the  intrinsic $A$-nucleon microscopic Hamiltonian, which it is useful to decompose into, e.g.:
\begin{equation}
H=T_{\rm rel}(r)+ {\mathcal V}_{\rm rel} +\bar{V}_{\rm C}(r)+H_{(A-a)}+H_{(a)}\,.
\end{equation}
Further, $T_{\rm rel}(r)$ is the relative kinetic energy and ${\mathcal V}_{\rm rel}$ is the sum of all interactions between nucleons belonging to different clusters after subtraction of the avarage Coulomb interaction between them, explicitly singled out in the term $\bar{V}_{\rm C}(r)=Z_{1\nu}Z_{2\nu}e^2/r$, where $Z_{1\nu}$ and $Z_{2\nu}$ are the charge numbers of the clusters in channel $\nu$. 
\begin{table*}[t]
\begin{ruledtabular}
\begin{tabular}{c c c c c c c c c c c c}
&&$^3$H&&\multicolumn{8}{c}{$n\,$-${}^3$H ($E_{\rm kin}=0.75$ MeV)}\\\cline{3-3}\cline{5-12}
$N_{\rm max}$&&$E_{\rm g.s.}$&&$0^+$ ($^1S_0$)&$0^-$ ($^3P_0$)&$1^+$ ($^3S_1$)&$1^-$ ($^1P_1$)& $1^-$ ($^3P_1$)&$1^-$ $(\epsilon)$&$2^+$ ($^3P_2$)&$\sigma_t$\\[0.7mm]
\hline
$9$&&$-7.80$&&$-27.8$&$2.30$&$-26.2$&$2.19$&$4.96$&$-17.5$&$7.51$&$1.06$\\
$11$&&$-7.96$&&$-31.3$&$2.39$&$-28.1$&$2.63$&$5.93$&$-12.7$&$6.42$&$1.20$\\
$13$&&$-8.02$&&$-32.4$&$2.15$&$-28.8$&$3.10$&$6.17$&$\,\,-9.1$&$5.75$&$1.25$\\
$15$&&$-8.11$&&$-33.2$&$2.45$&$-29.9$&$3.46$&$6.12$&$\,\,-9.5$&$6.08$&$1.33$\\
$17$&&$-8.12$&&$-34.2$&$2.60$&$-30.9$&$3.74$&$6.30$&$-10.7$&$6.19$&$1.41$\\
$19$&&$-8.16$&&$-34.8$&$2.49$&$-31.3$&$4.00$&$6.49$&$-10.1$&$6.02$&$1.44$\\[1mm]
&&$^4$He&&\multicolumn{4}{c}{$n\,$-${}^4$He ($E_{\rm kin}=5.0$ MeV)}&&\multicolumn{3}{c}{$p\,$-${}^4$He ($E_{\rm kin}=5.0$ MeV)}\\[0.4mm]\cline{3-3}\cline{5-8}\cline{10-12}
$N_{\rm max}$&&$E_{\rm g.s.}$&&$\frac{1}{2}^+$ ($^2S_{1/2}$)&$\frac{1}{2}^-$ ($^2P_{1/2}$)&$\frac{3}{2}^-$ ($^2P_{3/2}$)&$\sigma_t$&&$\frac{1}{2}^+$ ($^2S_{1/2}$)&$\frac{1}{2}^-$ ($^2P_{1/2}$)&$\frac{3}{2}^-$ ($^2P_{3/2}$)\\[0.9mm]
\hline
$9$&&$-27.00$&&$-57.9$&$33.5$&$81.8$&$1.95$&&$-45.8$&$31.3 $&$76.5$\\
$11$&&$-27.41$&&$-58.6$&$33.7$&$86.1$&$1.98$&&$-46.4$&$31.9$&$80.2$\\
$13$&&$-27.57$&&$-58.7$&$34.0$&$85.7$&$1.98$&&$ -46.6$&$32.0$&$80.0$\\
$15$&&$-27.75$&&$-58.7$&$33.9$&$84.6$&$1.97$&&$ -46.6$&$32.1$&$79.9$\\
$17$&&$-27.77$&&$-58.6$&$33.9$&$84.8$&$1.97$&&$ -46.5$&$32.0$&$79.9$
\end{tabular}
\end{ruledtabular}
 \caption{Calculated $^3$H and $^4$He g.s.\ energies (in MeV), $n\,$-${}^3$H, $n\,$-${}^4$He and $p\,$-${}^4$He phase shifts (in degrees), and $n\,$-${}^3$H and $n\,$-${}^4$He total cross sections (in barns)  for increasing $N_{\rm max}$ at $\hbar\Omega$ = $18$ MeV, obtained using the $V_{lowk}$ NN potential (derived from AV18 with cutoff $\Lambda=2.1$ fm${}^{-1}$)~\cite{BoKu03}. Only the g.s.\ of the ${}^3$H and ${}^{4}$He nuclei were included in the scattering calculations.}\label{tab}   
\end{table*}

We obtain the cluster eigenstates entering Eq.~(\ref{basis}) by diagonalizing $H_{(A-a)}$ and $H_{(a)}$ in the model space spanned by the NCSM basis. This is a complete harmonic oscillator (HO) basis, the size of which is defined by the maximum number, $N_{\rm max}$, of HO quanta above the lowest configuration shared by the nucleons. Thanks to the unique properties of the HO basis,  we can make use of Jacobi-coordinate wave functions~\cite{Jacobi_NCSM} for both nuclei or only for the lightest of the pair (typically $a\le4$), and still preserve the translational invariance of the problem. In the second case we expand the heavier cluster on a Slater-determinant (SD) basis, and remove completely the spurious c.m.\ components in a similar fashion as in Refs.~\cite{tr_dens, cluster}.  We exploited this dual approach to verify our results.  The use of the SD basis is computationally advantageous and allows us to explore reactions involving $p$-shell nuclei. In calculating~(\ref{K-kernel}), all ``direct'' terms arising from the identical permutations in both $\hat{\mathcal A}_{\nu}$ and $\hat{\mathcal A}_{\nu^\prime}$ are treated exactly with the exception of  $\left\langle\Phi^{J^\pi T}_{\nu^\prime r^\prime}\right|{\mathcal V}_{\rm rel}\left|\Phi^{J^\pi T}_{\nu r}\right\rangle$. The latter and all remaining terms are obtained by expanding the Dirac $\delta$ of Eq.~(\ref{basis}) on a set of HO radial wave functions with identical frequency $\Omega$, and model-space size 
$N_{\rm max}$ consistent with those used for the two clusters. In this respect we note that ${\mathcal V}_{\rm rel}$ is localized also in presence of the Coulomb force. 
We solve Eq.~(\ref{RGMeq}) by means of the coupled-channel $R$-matrix method on a Lagrange mesh~\cite{R-matrix} imposing either bound-state or scattering boundary conditions for $g^{J^\pi T}_{\nu}(r)$  at large $r$.

All calculations in the present paper were carried out using binary-cluster channels~(\ref{basis}) with $a$ = 1. We first discuss results obtained limiting the expansion~(\ref{trial}) to configurations with the $(A-1)$-cluster in its g.s. Table~\ref{tab} shows the behavior with respect to $N_{\rm max}$ of selected $A=4,5$ data obtained using the $V_{lowk}$ NN potential~\cite{BoKu03}. 
A satisfactory convergence of both g.s.\ energies and scattering data is reached starting from $N_{\rm max}=17$,  for the four-, and $N_{\rm max}=15$ for the five-nucleon systems, respectively.

In what follows we present results obtained using effective interactions derived from the underlying realistic NN potential, $V_N$, through a unitary transformation.
\begin{figure}[b]
\includegraphics*[scale=0.545]{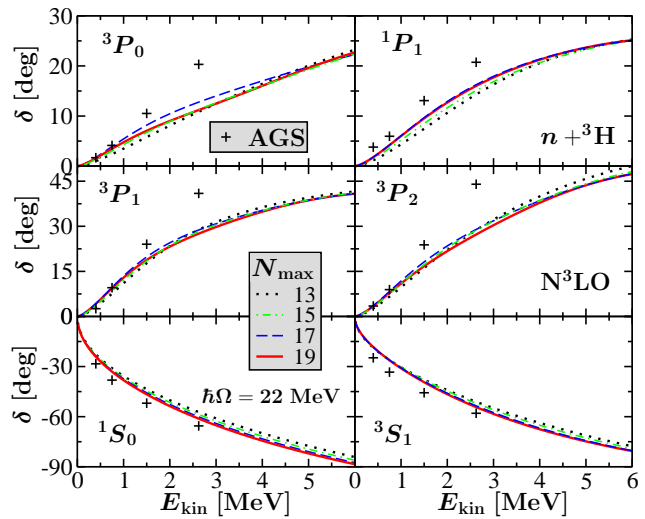}%
\caption{(Color online.) Calculated phase shifts for $n\,$-${}^3$H scattering as a function of the relative kinetic energy in the c.m.\ frame $E_{\rm kin}$, using the N$^3$LO NN potential~\cite{N3LO}. Only the g.s.\ of the $(A-1)$-cluster was included in the present calculation.
Dependence on the model-space truncation $N_{\rm max}$ at $\hbar\Omega=22$ MeV compared to AGS results of Ref.~\cite{Deltuva, DeltuvaPriv}.}\label{n3H}
\end{figure}
Starting from the relevant  two-nucleon Hamiltonian (for notation and definitions  see Ref.~\cite{Jacobi_NCSM}) $H^{\Omega}_{2} = H_{02}+V_{12}$, with $V_{12}=V_{N}(\sqrt{2}\vec{r}\,) - m\Omega^2\vec{r}^{\,2}/A$, the cluster eigenstates are obtained employing the usual NCSM two-body effective interaction $V_{2\rm{eff}}=\bar{H}_{2\rm{eff}}-H_{02}$, where $\bar{H}_{2\rm{eff}}$ is the Hermitean effective Hamiltonian. 
However, in place of the bare NN potential entering ${\mathcal V}_{\rm rel}$ we adopted the new effective interaction $V^\prime_{2\rm{eff}}=\bar{H}_{2\rm{eff}}-\bar{H}^\prime_{2\rm{eff}}$, where $\bar{H}^\prime_{2\rm{eff}}$ is the effective Hamiltonian derived from $H^{\Omega\,\prime}_2 = H_{02}+V^\prime_{12}$, with $V^\prime_{12} = - m\Omega^2\vec{r}^{\,2}/A$. 
Note that $V^\prime_{2\rm{eff}}\rightarrow V_{N}$ in the limit $N_{\rm max}\rightarrow\infty$. 
\begin{figure}[t]
\includegraphics*[scale=0.65]{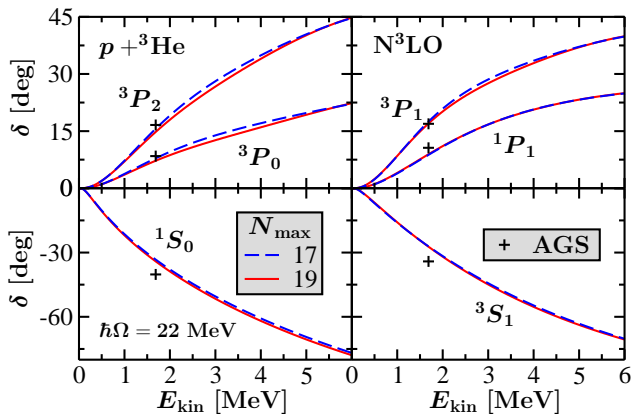}%
\caption{(Color online.)  Same as Fig.~\ref{n3H} for $p$ -$^3$He scattering.}\label{p3He}
\end{figure}
Figures~\ref{n3H},~\ref{p3He} and the right panel of Fig.~\ref{na} present $A=4,5$ scattering phase shifts for the high-quality NN potential derived within chiral effective-field theory at next-to-next-to-next-to-leading order (N$^3$LO)~\cite{N3LO}. For the whole energy range, we find   
less than $2$ deg absolute difference between the phases obtained in the largest and next-to-largest model spaces, a sign of convergence. The only exception is represented by the $^2P_{3/2}$  phase shifts of the $n$-$\alpha$ system, for which this difference rises up to $5$ deg  in the
range $1$ MeV$<E_{\rm kin}<4$ MeV. As a comparison, we show in the left panel of Fig.~\ref{na} the $n$-$\alpha$ phase shifts obtained with the (bare) $V_{lowk}$ interaction. The convergence rate is clearly much faster.  

In order to verify our approach, in Fig.~\ref{n3H} and~\ref{p3He} we compare our $n\,$-${}^3$H and $p\,$-${}^3$He results to earlier {\em ab initio} calculations performed in the framework of the Alt, Grassberger and Sandhas (AGS) equations~\cite{Deltuva,DeltuvaPriv}, using the same N$^3$LO NN potential. We note that in general  the agreement between the two calculations worsens as the relative kinetic energy in the c.m.\ frame, $E_{\rm kin}$,  increases. For the $P$-waves in particular we can reasonably reproduce the AGS calculation  for energies within $1$ MeV off threshold, while we can find differences as large as $17$ deg ($^3P_2$) at $E_{\rm kin}=2.6$ MeV. 
These discrepancies are due to the influence, increasing with energy, played by closed channels not included in our calculations, such as those with the $A-1=3$ eigenstates above the $I_1^{\pi_1}=\frac{1}{2}^+$ g.s., and ($A-a=2,a=2$) configurations, present in the AGS results. 
\begin{figure}[t]
\includegraphics*[scale=0.65]{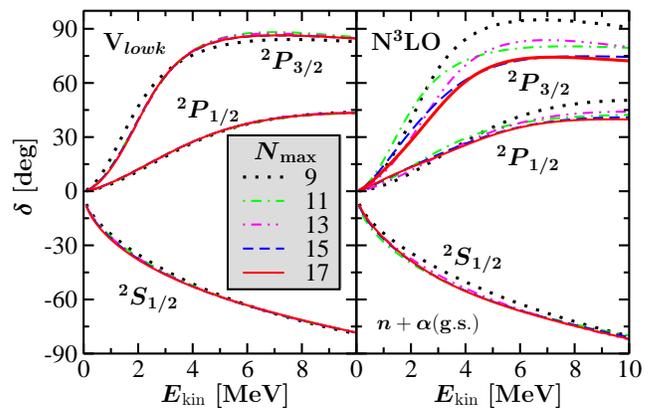}%
\caption{(Color online.) Same as Fig.~\ref{n3H} for $n$-$\alpha$ scattering with $V_{lowk}$ NN potential~\cite{BoKu03} at $\hbar\Omega=18$ MeV (left panel), and N$^3$LO NN potential~\cite{N3LO} at $\hbar\Omega=19$ MeV (right panel).}\label{na}
\end{figure}
In Ref.~\cite{Deltuva} it was shown that the omission of three-nucleon partial waves with $\frac{1}{2}<I_1\le\frac{5}{2}$ leads to effects of comparable magnitude on the AGS results at $E_{\rm kin}=3$ MeV, especially for the $^3S_1, ^3P_1$ and $^3P_2$. 
\begin{figure}[b]
\includegraphics*[scale=0.65]{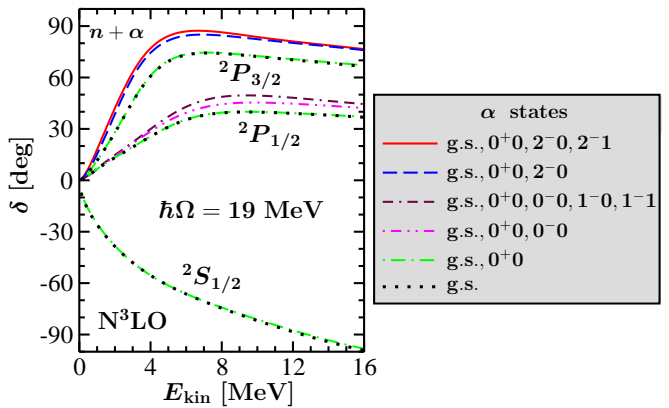}%
\caption{(Color online.) Influence of  the lowest six exited states ($0^+ 0, 0^- 0,1^-0,1^-1, 2^- 0,2^-1$) of the $\alpha$ particle on the $n$-$\alpha$ phase-shift results for the N$^3$LO NN potential~\cite{N3LO}.} \label{napol}
\end{figure}

We explore the effect of the inclusion of higher excited states of the $(A-1)$-cluster on the $n$-$\alpha$ scattering phase shifts. Channels with $a>1$ have  here a much suppressed effect due to the large binding energy of the $^4$He nucleus. Figure~\ref{napol} shows the influence of the six lowest excited states of $^4$He on the $n$-$\alpha$ phase shifts. The $I_1^{\pi_1}T_1=0^+ 0$ excited state affects only minimally the $^2S_{1/2}$, leaving the $P$ phase shifts unaltered. On the contrary, we find larger deviations on the $^2P_{1/2}$ and $^2P_{3/2}$ phase shifts, after the inclusion of the $0^- 0, 1^-0,$ and $1^-1$ states for the first, and of the $2^- 0$ and $2^-1$ states for the second. These negative-parity states do not influence the $^2S_{1/2}$. 
\begin{figure}[t]
\includegraphics*[scale=0.65]{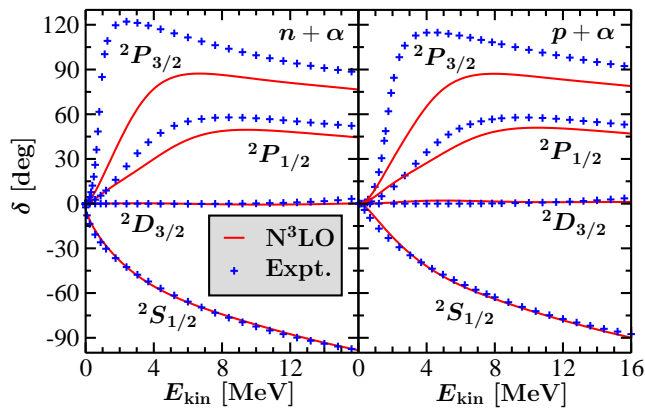}%
\caption{(Color online.) Calculated $n$-$\alpha$ (left panel) and $p\,$-$\alpha$ (right panel) phase shifts for the N$^3$LO NN potential~\cite{N3LO}, including the $^4$He g.s., $0^+ 0$, $0^- 0$, $1^-0$, $1^-1$,  $2^- 0$, and $2^-1$ states, compared to an $R$-matrix analysis of data ($+$)~\cite{HalePriv}.}\label{npaexp}
\end{figure}
\begin{table}[b]
\begin{ruledtabular}
\begin{tabular}{lccclrclr}
&&$^{10}$Be&&\multicolumn{2}{c}{$^{11}$Be($\frac12 ^-$)}&&\multicolumn{2}{c}{$^{11}$Be($\frac12 ^+$)}\\[0.7mm]\cline{3-3}\cline{5-6}\cline{8-9}\\[-4mm]
&$N_{\rm max}$&$E_{\rm g.s.}$ &&\multicolumn{1}{c}{$E$}&\multicolumn{1}{c}{$E_{th}$}&&\multicolumn{1}{c}{$E$}&\multicolumn{1}{c}{$E_{th}$}\\[0.5mm]
\hline
NCSM~\cite{10Be,11Be} & 8/9& -57.06&&-56.95&0.11&&-54.26&2.80\\
NCSM~\cite{10Be,11Be},$^{a}$ & 6/7& -57.17&&-57.51&-0.34&&-54.39&2.78\\
NCSM/RGM\footnote{present calculation} &&&&-57.59&-0.42&&-57.85&-0.68\\
Expt. & & -64.98&&-65.16&-0.18&&-65.48&-0.50 
\end{tabular}
\end{ruledtabular}
\caption{Calculated energies (in MeV) of the $^{10}$Be g.s.\ and  of the lowest negative- and positive-parity states in $^{11}$Be, obtained using the CD-Bonn NN potential~\cite{CD-Bonn2000} at $\hbar\Omega=13$ MeV. The NCSM/RGM results were obtained using $n+^{10}$Be configurations with $N_{\rm max}$ = 6 g.s., $2^+_1$, $2^+_2$, and $1^+_1$ states of $^{10}$Be.}\label{11Be}
\end{table}

In Fig.~\ref{npaexp} the $n$- and $p\,$-$\alpha$ phase shifts obtained with the N$^3$LO NN potential, including the first six $^4$He excited states, are compared to the results of an accurate multi-channel $R$-matrix analysis of the nucleon-$\alpha$ scattering data~\cite{HalePriv}. The $^2S_{1/2}$ phase shifts are in good agreement with experiment, also in presence of the Coulomb repulsion between proton and $\alpha$ particle.
The magnitude of the $^2D_{3/2}$ phase shifts is also qualitatively reproduced. On the contrary, the $P$ phase shifts present both insufficient magnitude and splitting with respect to the predictions of the $R$-matrix analysis. The $\frac12^+$ channel is dominated by the repulsion between nucleon and $\alpha$ particle induced by the Pauli exclusion principle. Consequently, the short-range details of the nuclear interaction play a minor role on the $^2S_{1/2}$ phase shifts, for which, as shown in Fig.~\ref{na}, we find very similar results using the $V_{lowk}$ potential. On the other hand, the latter figure shows also that the $^2P_{1/2}$ and $^2P_{3/2}$ phase shifts are sensitive to the interaction model, and in particular to the strength of the spin-orbit force. The present discrepancy with respect to experiment is due to the omission of the three-nucleon terms of the chiral interaction at order N$^3$LO, which would lead to an enhanced spin-orbit splitting.
\begin{figure}[t]
\includegraphics*[scale=0.65]{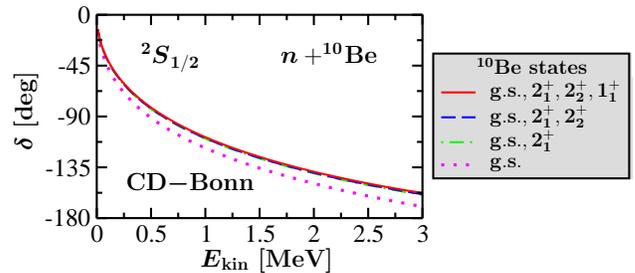}%
\caption{(Color online.) Calculated $^2S_{1/2}$ $n\,$-${}^{10}$Be  phase shifts as a function of $E_{\rm kin}$, using the CD-Bonn NN potential~\cite{CD-Bonn2000} at $\hbar\Omega=13$ MeV. NCSM/RGM calculation as in Tab.~\ref{11Be}.}\label{n10Be}
\end{figure}

To show the promise and flexibility of our approach,  we present in Tab.~\ref{11Be} and Fig.~\ref{n10Be} results for a much heavier ($A$ = 11) system. The parity-inverted g.s.\ of $^{11}$Be, one of the best examples of disappearance of the $N$ = 8 magic number with increasing $N/Z$ ratio, was so far left unexplained by  {\em ab initio} calculations~\cite{11Be}. The HO asymptotic behavior of the $^{11}$Be wave function in the standard NCSM does not favor extended $n\,$-${}^{10}$Be configurations, thus enhancing the relative kinetic energy repulsion, and preventing the experimentally-observed inversion between $\frac12 ^-$ and $\frac12 ^+$ states. Using the CD-Bonn NN potential~\cite{CD-Bonn2000}, we observe a dramatic ($\sim3.5$ MeV) increase in the $^{11}$Be $\frac12 ^+$ state binding energy leading to the g.s.\ parity inversion, when the $n\,$-${}^{10}$Be relative motion is treated within the {\em ab initio} NCSM/RGM approach.

We thank A.\ Deltuva, P.\ Descouvemont, J.\ Hale, and I.\ J.\ Thompson for valuable discussions.
Numerical calculations have been performed at the LLNL LC facilities.
Prepared by LLNL under Contract DE-AC52-07NA27344.
Support from the U.S.\ DOE/SC/NP (Work Proposal No.\ SCW0498), 
and from the U.\ S.\ Department of Energy Grant DE-FG02-87ER40371 is acknowledged.

\end{document}